\documentclass{article}
\usepackage{graphicx}
\usepackage{subfig}
\usepackage{xcolor}
\usepackage[noadjust]{cite}

\date{}

\author{
L. Bastianelli$^{1,2}$, R. Diamanti$^3$, E. Colella$^{1,2}$, V. Mariani Primiani$^{1,2}$, \\ F. Moglie$^{1,2}$,  M. A. Toubal$^4$, M. Odit$^4$, J.-B. Gros$^4$, Y. Nasser$^4$, G. Lerosey$^4$, \\ L. Santamaria$^4$, A. Allasia$^5$, M. Crozzoli$^5$, M. Boldi$^5$, \\ E. Zimaglia$^5$, V. Lieti$^6$, M. Colombo$^6$, D. Micheli$^7$ \\ \\
1 Universit\`a Politecnica delle Marche, Ancona, Italy \\ 
2 Consorzio Nazionale Interuniversitario per le Telecomunicazioni, Parma, Italy \\
3 TIM S.p.A., Ancona, Italy  \\
4 Greenerwave, Paris, France \\ 
5 TIM S.p.A., Turin, Italy \\ 
6 Nokia Networks Italia, Vimercate, Italy \\
7 TIM S.p.A., Rome, Italy \\
}

\begin{document}

\title{Measurements of Reconfigurable Intelligent Surface in 5G System Within a Reverberation Chamber at mmWave}

\maketitle

\begin{abstract}
In this paper we evaluated the performance of a reconfigurable intelligent surface tested with a fifth generation signal provided by a commercial
fifth generation base station. We adopted a reverberation chamber as a real life propagating environment.
Tests were conducted at the millimeter wave frequency range.
This measurement campaign was carried out under the H2020 European project RISE-6G and a
collaboration program between TIM S.p.A., Nokia and Universit\`a Politecnica delle Marche.
\end{abstract}

\vskip0.5\baselineskip
%\begin{IEEEkeywords}
%Reverberation chamber, 5G, base station, RIS, mmWave, live radio access network, RSRP.
%\end{IEEEkeywords}

\section{Introduction}

The fifth generation (5G) technology is in the stage of deployment and the doors are already opening for the 6G~\cite{DiRenzo-road6G}.
In spite of the new services and opportunities provided by the 5G, there are still many challenges.
In the 5G networks are present massive multiple-input multiple-output (MIMO) implementations in antenna design, in some case with
more than one hundred elements, and the allocation of the millimeter wave (mmWave)
and sub-THz spectrum beside the already sub-6 GHz spectrum employed by the mobile operators.
The upcoming communication networks are expected to comply with extremely demanding requirements that are not always easily achievable, such as
expecting to serve a very large number of users and devices, a high data rates and ensuring a very low
latency through ultra-dependable low-latency communications (URLLC).
Another key aspect is the energy efficiency given the large number of devices and base stations that will be employed.
By increasing the available spectrum, with mmWave and sub-THz frequency ranges~\cite{sakaguchi,Choudhury,kurner}, the 5G and the upcoming 6G technology will drive
a significant increase in mobile data traffic and machine-to-machine (M2M) communications and internet-of-things (IoT).
This new spectrum guarantees both high interconnection between nodes and devices and a high-speed mobile data transmission.
At these frequencies, in a real propagation scenario the electromagnetic waves that encounter obstacles during propagation are subject to significant signal loss.
In some scenarios, these obstacles may remain fixed like a building but in other cases they may be moving.
This is a challenge that can be mitigated by adopting some strategies like implement massive MIMO solutions and
by using smart antennas, able to overcome multipath fading, thus ensuring an acceptable communication link~\cite{Alexandropoulos1}.
In recent years, emerging technologies such as reconfigurable intelligent surfaces (RISs) are the subject of
numerous studies and research~\cite{rise-paper1,rise-paper2,DiRenzo-RIS,RIS-sub6,rise6}.
The RIS is basically a very thin sheet capable of modifying the electromagnetic radio waves impinging upon it~\cite{RIS1,Geoffroy-RIS,Geoffroy-RIS2,V-Esposti}.
They can be programmed to pursue a specific operation and may also be able to reconfigure themselves after installation in the environment.
The RISs can be integrated in the network architecture in order to make the whole radio environment ''smart'' thus saving energy and reduce the installation
of base stations~\cite{RIS-energy}. With these premises the RISs will have a relevant role in upcoming wireless systems.
The integration of RISs with 5G systems presents a number of challenges like electromagnetic factors, protocols to communicate with
the existing networks, localization and so on.
The measurement campaign reported in this paper show the capability of the RIS to focus the received signal from the 5G
base station (BS) antenna to the receiver. 
The over-the-air (OtA) tests are performed by using a reverberation chamber (RC).
The RC allows us to replicate real-life complex propagation conditions for both the BS and user equipment (UE) and we can perform measurements 
in a controlled way in our laboratory by avoiding external interferences.
The peculiarity of the RC of emulating indoor/outdoor scenarios makes it very useful for this type of measurements.
In fact, it is widely used for testing wireless devices~\cite{micheli2,micheli3,micheli4,holloway2}.
This work is a collaboration between CNIT-Unit research of Ancona, TIM and Greenerwave all of them involved in the
RISE-6G project, and Nokia as BS supplier in these tests.

\section{Measurements set-up and results}

\begin{figure}[h]
\centering
\includegraphics[width=0.30\columnwidth]{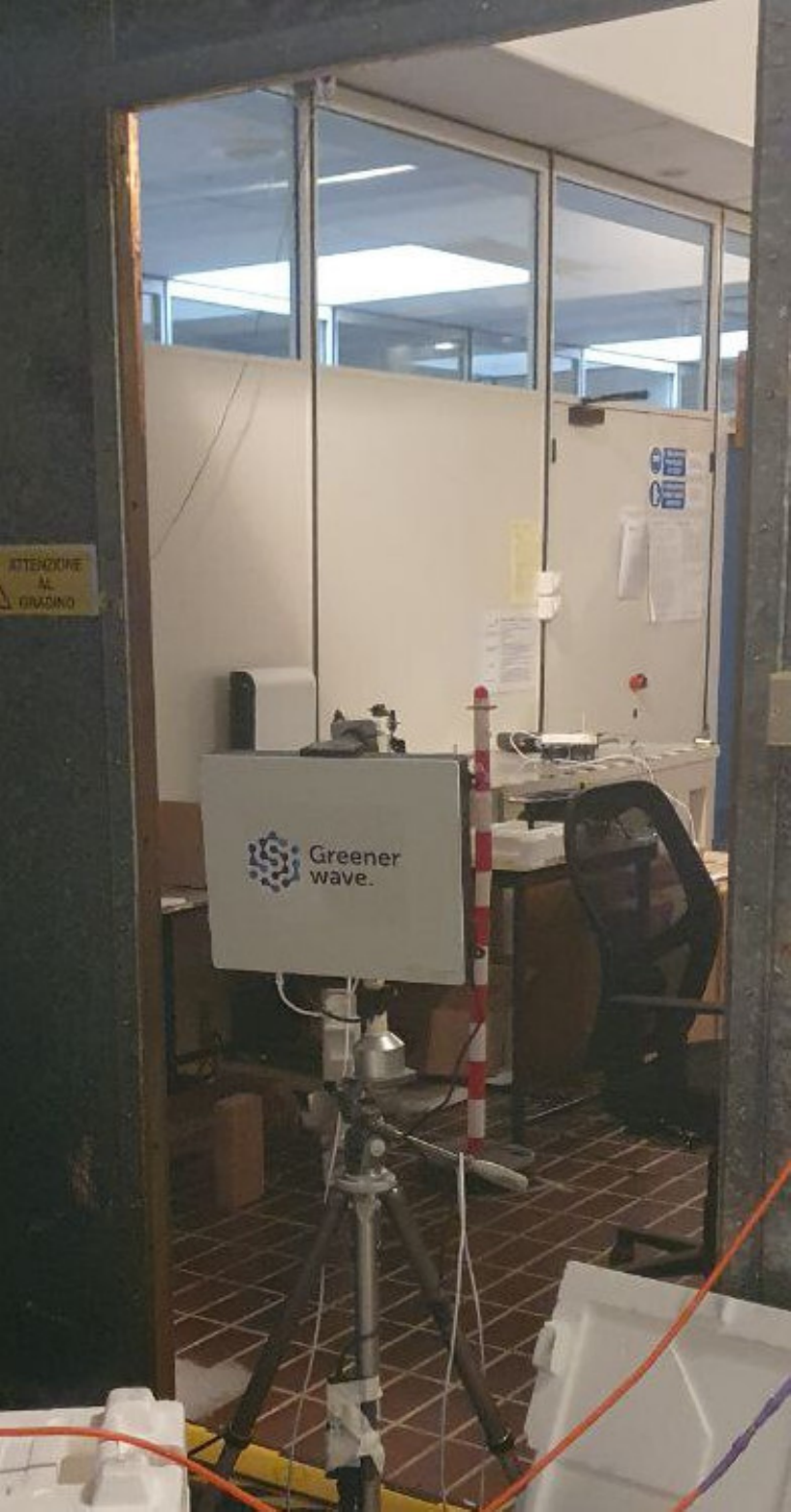}
\caption{Picture of the RIS. It is placed behind the RC's door.}
\label{fig:ris_door}
\end{figure}

\begin{figure}[h]
\centering
\includegraphics[width=0.45\columnwidth]{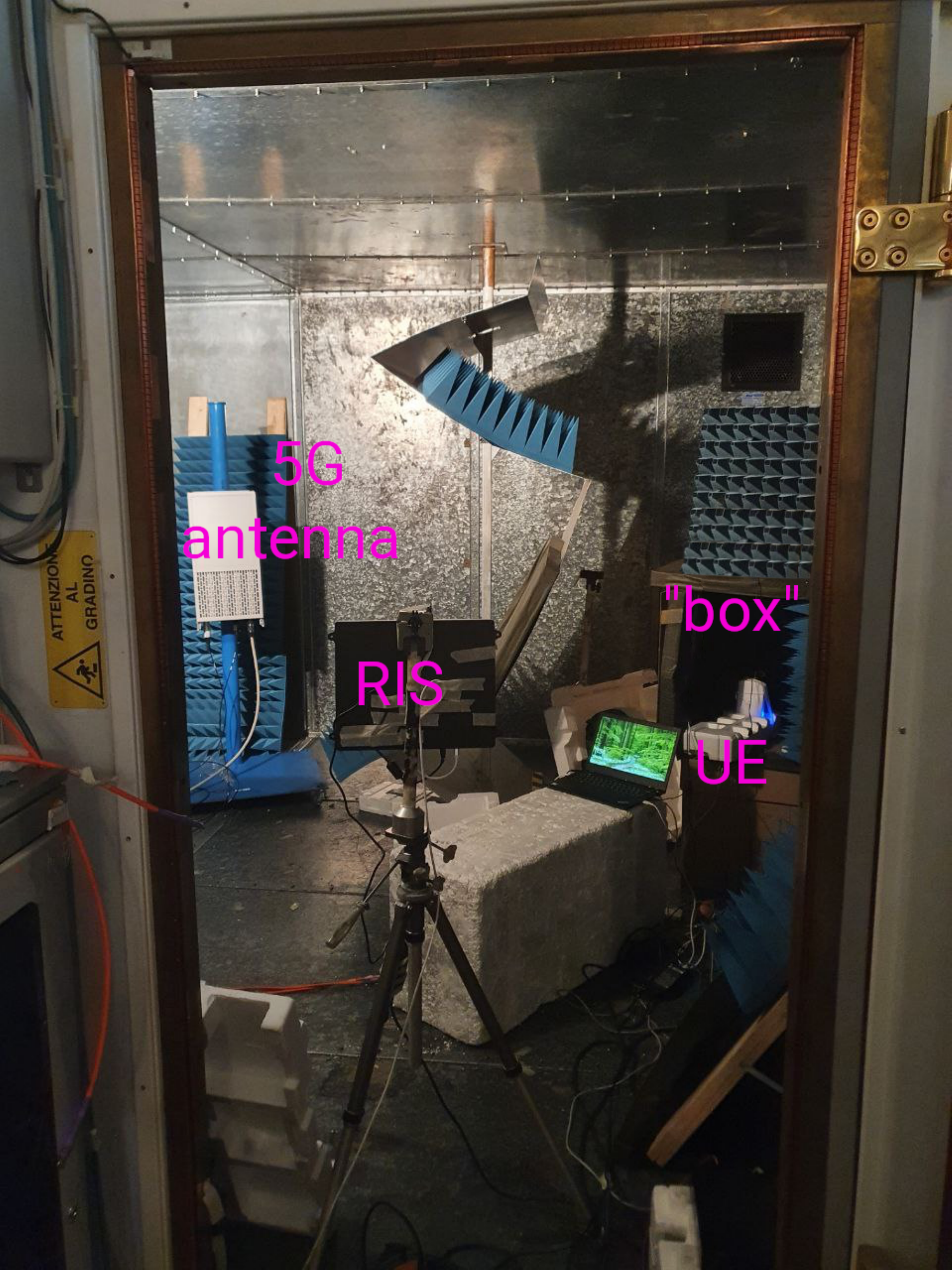}
\caption{Picture of the set-up. 5G base station antenna, RIS, UE and absorbing material are within the RC.}
\label{fig:setup}
\end{figure}

In this measurement campaign we used the RC to emulate a multipath environment for testing wireless devices.
The adoption of the RC for this kind of tests is well known for who provides the base station (BS), wireless devices and also for mobile
operators~\cite{remleyRC, magazine, skarbratt2011, andersson}.
Different loading conditions present variations in the characteristic parameters of the RC, such as: the quality factor ($Q$-factor), the power delay profile (PDP), the time delay spread ($\tau_{RMS}$) and the $K$-factor.
The RC is able to reproduce a desired real-life environments~\cite{micheli1,micheli0,micheli0a}.
The ability to use the RC in the laboratory allows us to emulate different propagation scenarios by changing the load inside it instead of moving instruments and people in multiple environments.
By varying the amount and position of absorbing materials it is possible to emulate the chosen environment; it can be checked
by matching the $\tau_{RMS}$ according to international standards~\cite{guide1,guide2,guide3}.
The more the chamber is loaded, the lower the $Q$-factor. As consequence, in a loaded chamber the PDP exhibits a more pronounced decay and also the $\tau_{RMS}$ decreases.
The procedure for the evaluation of those parameters in different loading conditions is reported in the literature~\cite{holloway,remley,dresda}.
In~\cite{eucap23} has been conducted a preliminary activity at 27~GHz in the RC with: i) a commercial 5G BS connected to the TIM live network;
ii) a 5G smart antenna with analog beamforming; iii) an user equipment (UE) to collect data; iv) vertical and horizontal stirrers; v) absorbing materials;
vi) a personal computer (PC).
With respect to~\cite{eucap23}, in these measurements we added an RIS within the RC. Figure~\ref{fig:ris_door} shows the RIS that is placed behind the door of the RC.
The measurement set-up is reported in~\figurename~\ref{fig:setup}. 
The set-up was arranged by ensuring: i) a line-of-sight (LoS) path between the transmitting antenna and the RIS; ii) a LoS path between the RIS and the UE.
The LoS path between the antenna and the UE was blocked by the vertical stirrer which was also covered by absorbers.
Moreover, the UE was placed within a ``box'' covered by absorbers.
The absorbing box reduces the multipath contributes due to metallic walls and it also contributes to emphasize the effect due to the RIS.
The operating frequency of the BS and the RIS is 27~GHz, which corresponds the licensed frequency.

The mmWave RIS has been designed and provided by Greenerwave~\cite{GNW-RIS}, see~\figurename~\ref{fig:ris_door}.
It can operate in the frequency band from 27~GHz to 31~GHz.
\begin{figure*}[h]
\centering
\subfloat[Horizontal radiation pattern.]{\includegraphics[width=0.65\columnwidth]{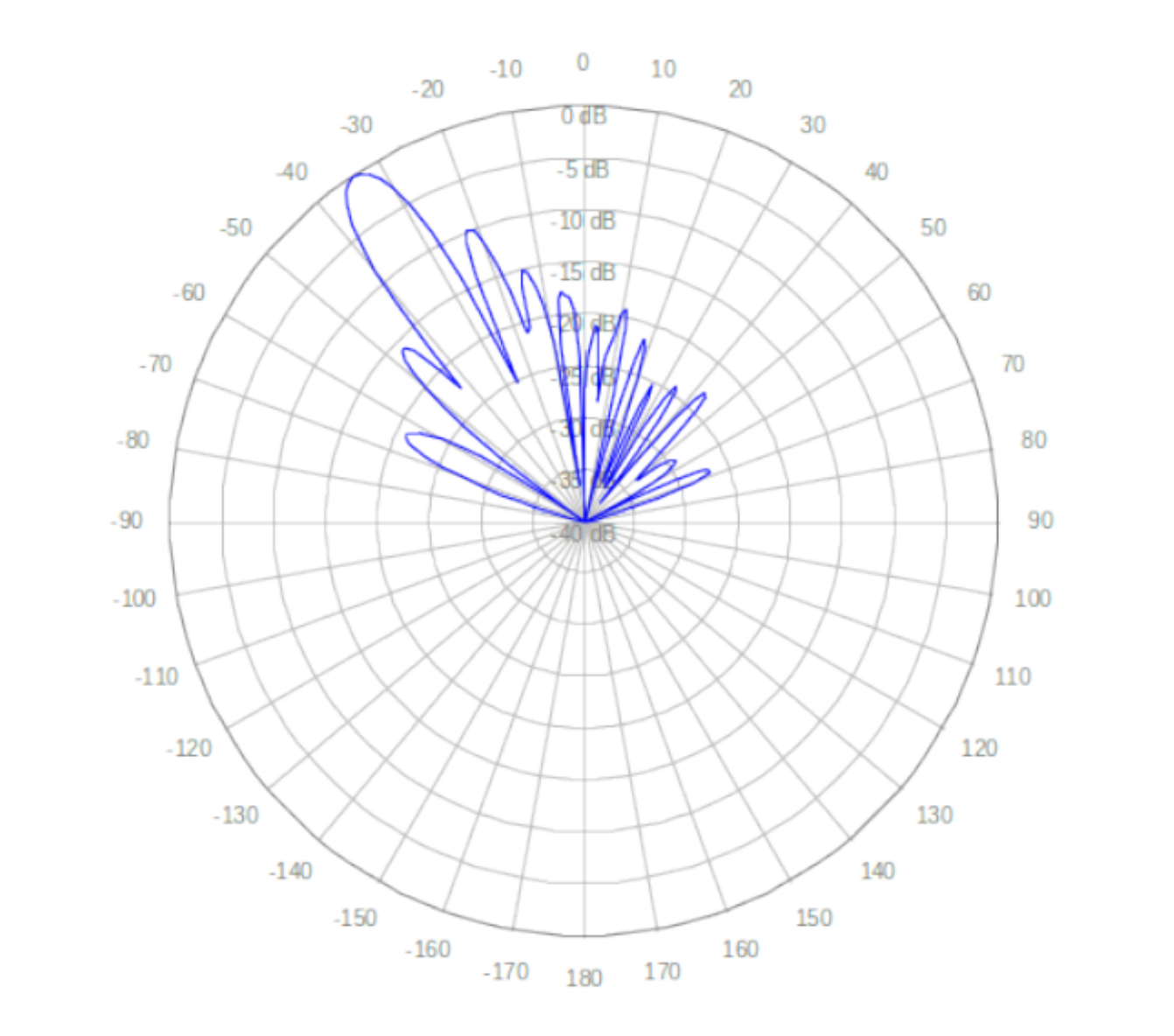}
\label{fig:hor_pat}}
\hfil
\subfloat[Vertical radiation pattern.]{\includegraphics[width=0.65\columnwidth]{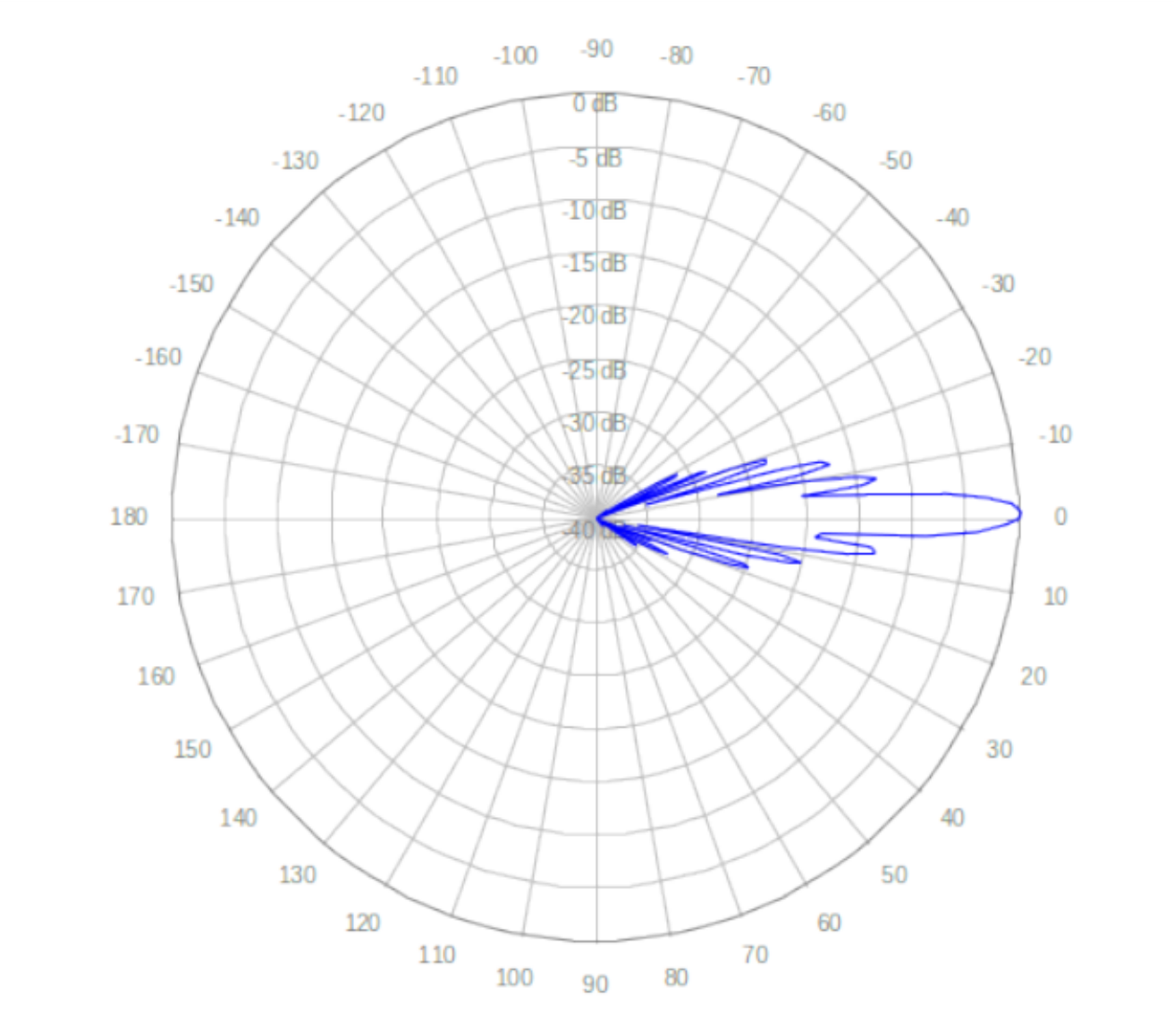}
\label{fig:ver_pat}}
\caption{Radiation pattern of the selected beam during measurements.}
\label{fig:rad_pat}
\end{figure*}
The switching between the resonance states is performed by means the control of the ON/OFF states of PIN diodes.
In this band, the phase difference between the two states ON and OFF is 180~deg. The RIS consists of 3200~diodes.
The PC controls the RIS via software by selecting the ON/OFF states of the diodes to focus the signal on a specific location.
Moreover, the PC is also connected to the UE in order to get data by using a proprietary software, i.e. the QXDM.

In these measurements we maintained fixed the beam of the antenna BS. 
The previous activity~\cite{eucap23} has been very useful.
On the one hand for the evaluation of the key performance indicators (KPIs)
of a commercial 5G BS operating at the mmWave frequency range (at 27~GHz), e.g. the modulation coding scheme (MCS), RSRP.
On the other hand for understanding the capability of the BS to switch beams, since the Nokia BS in our laboratory has 32 beams.
In order to evaluate the contribution of the RIS~\cite{Gros}, it is necessary to fix the transmitting beam pointing toward the RIS.
Otherwise, if we leave the BS free to choose the beam it might happen that the selected beam is not the one that points to the RIS, nullifying its effect.
Figure~\ref{fig:rad_pat} shows the radiation pattern of the selected beam. The radiation pattern of~\figurename~\ref{fig:rad_pat} is given by a tool included in the
BS software. In particular,~\figurename~\ref{fig:hor_pat} shows the horizontal radiation pattern
whereas~\figurename~\ref{fig:ver_pat} reports the vertical radiation pattern when we select and fix the beam \#3 of the 32 available~\cite{eucap23}.
The chosen beam in our tests has the azimuth angle is $-35$~deg whereas the elevation angle is $0$~deg.
This configuration allows us to point the transmission in LoS with the RIS.
\begin{figure}[h]
\centering
\includegraphics[width=0.85\columnwidth]{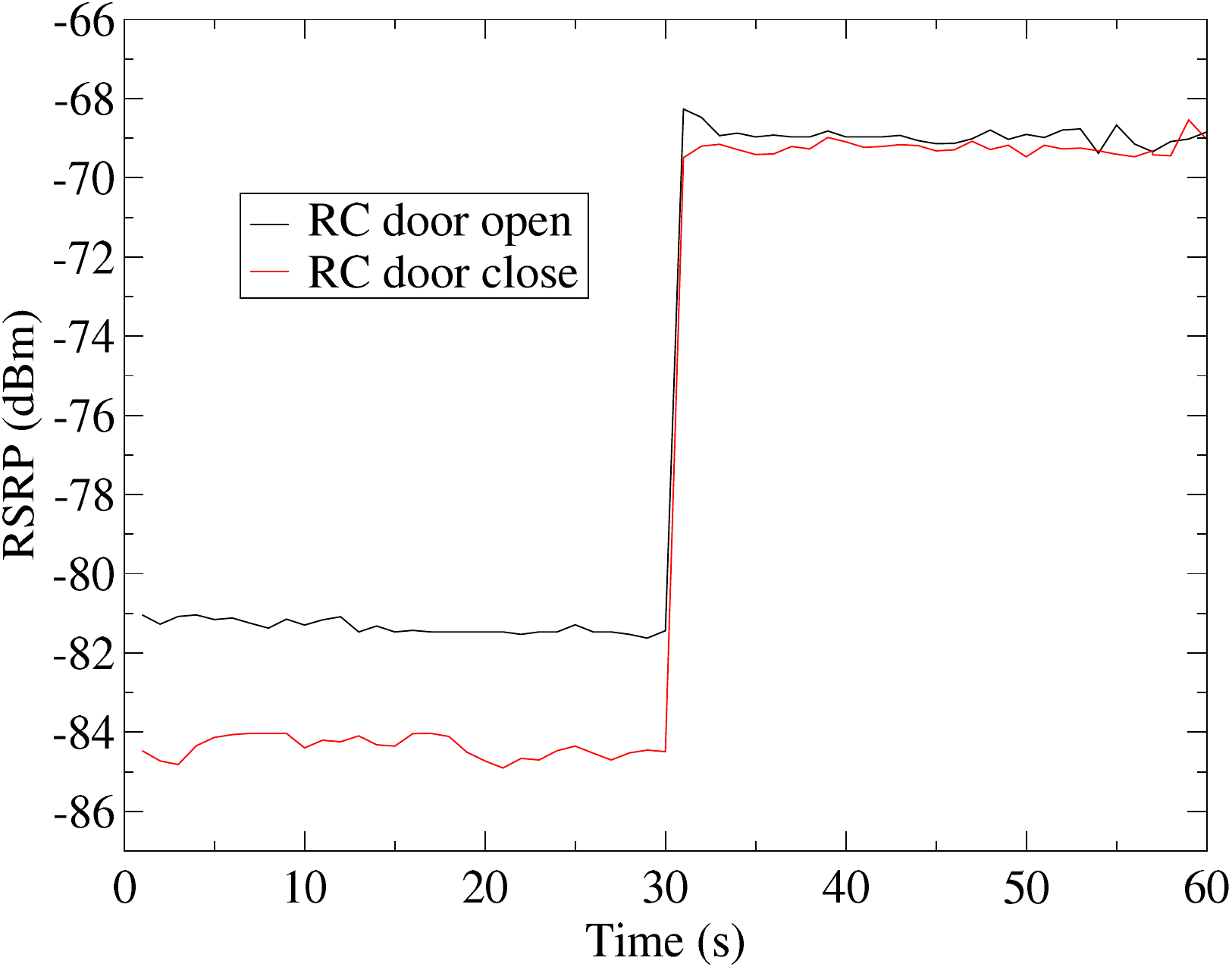}
\caption{RSRP recorded by the UE when the RIS initially has a random configuration and then, after 30~s, with a correct configuration.}
\label{fig:rsrp}
\end{figure}

Figure~\ref{fig:rsrp} shows the reference signals received power (RSRP) recorded by the UE.
During the acquisition we maintained active data traffic to the UE side.
In these tests we explored two configurations of the RIS:
\begin{itemize}
\item initially we loaded a random phase configuration to the RIS
\item the random configuration remains for 30 seconds
\item after 30 seconds, we loaded the ``best'' configuration to focus the 5G signal to the receiver.
\end{itemize}

Since the RIS is placed behind the RC door, one might expect that the RIS contribution is affected by the presence of this metallic wall that acts like a reflector.
To this purpose we investigated the behavior of the RIS when the door of the RC was open and closed.
By exploring the sequence of actions, RIS randomly configured and then opportunely configured, we noticed that the RIS has a relevant impact on the
link connection between the 5G antenna BS and the UE.
The RIS increased the RSRP level of about $12$~dB when the door was open whereas about $14$~dB when the door was closed.
The different level of the initial RSRP can be ascribed because the RIS has in both cases a random configuration that can affect the signal power level.
When the RIS was properly configured the RSRP is the same in both cases.
We can thus conclude that the metal wall behind the RIS does not affect its performance.

\section{Conclusion}

In this measurement campaign we evaluated the performance of an RIS at the mmWave frequency range.
We used the RC for tests because it allows us to reproduce real-life environments and stress the
operating conditions of the devices under test, namely the RIS.
For these tests we blocked the LoS path between the transmitting antenna and the UE but at the same time we
ensured the LoS path between: i) the transmitting antenna and the RIS; ii) the RIS and the UE.
Despite the high directivity of millimeter wave, the metal walls around the RIS, especially the door of the RC, could nullify its action.
By considering both the door open and closed we proved that the RIS is able to focus the received signal to receiver also in this more hostile condition.

\section*{Acknowledgment}

This work has been supported by EU H2020 RISE-6G project under the grant number 101017011.

\end{document}